\documentstyle[12pt,fleqn]{article}
\begin{document}
\title{ Diversity, Stability, Recursivity,  
and Rule Generation in Biological System:
Intra-inter Dynamics Approach}

\author{Kunihiko Kaneko\\
{\small \sl Department of Pure and Applied Sciences,
College of Arts and Sciences,}\\
{\small \sl University of Tokyo,}\\
{\small \sl Komaba, Meguro-ku, Tokyo 153, Japan}\\
}

\date{}

\maketitle

\begin{abstract}

Basic problems for the construction of a scenario for the Life are discussed.  
To study the problems in terms of dynamical systems theory,
a scheme of intra-inter dynamics is presented.
It consists of
internal dynamics of a unit, interaction among the units,
and the dynamics to change the dynamics itself, for
example by replication (and death) of units according to their internal
states.  Applying the dynamics to cell differentiation, 
isologous diversification theory is proposed.  According to it,
orbital instability leads to diversified cell behaviors first. At the next
stage, several cell types are formed, first
triggered  by clustering of oscillations, and then as
attracting states of internal dynamics stabilized by the cell-to-cell
interaction.  At the third stage, the differentiation is determined as a
recursive state by cell division.  At the last stage, hierarchical 
differentiation proceeds,
with the emergence of stochastic rule for the differentiation to sub-groups,
where regulation of the probability for the differentiation
provides the diversity and stability of cell society.  
Relevance of the theory to cell biology is discussed.
\end{abstract}

\section{Introduction: Life as Complex Systems}

A biological system generally
consists of diverse elements, which, as a total, has ability of reproduction.
In other words, a set of elements should reproduce itself,
although the reproduction must be rather loose and inaccurate,
when one imagines a prototype of Life.
To have such reproduction process efficiently, it is expected that
there are strong interactions among elements, since the reproduction
should involve positive feedback process,
which leads to nonlinear dynamics with instability in orbits.
As has been studied in nonlinear dynamics with many degrees of
freedom, the diversity of element states is then expected.  In this
sense one can expect that the diversity exists 
at the first stage of Life.  In general, (i) {\sl mechanism to create the
diversity} is the first important question to be
answered in the study of theoretical biology.
These diversified elements form an ensemble, which often keeps
some sort of stability, as an ensemble.  Organization of such 
higher-level is seen in a multi-cellular organism and  ecosystem.
(ii){\sl Mechanism of the formation of a higher (ensemble) level keeping 
stability} is the second question to be answered.

However, diversity and complexity are not same.  As for complexity,
some semantic structure is necessary.  In other words,  the
emergence of rules leading to a structure is required, and furthermore, 
subjective individual to distinguish the world as
complex rather than random is postulated \cite{KK-IKE}.
Then the following two questions are raised;
(iii) {\sl how a rule is formed from the mess of mutual
relationships}, and (iv) {\sl how an individual unit is formed to 
separate it out from the remaining world}.  (Possibly these two are related).

These four basic questions lead to more specific ones.
In spite of the tendency forming diversity,
(v){\sl how is a set of discrete types formed}?  Cells in a multi-cellular
organism are grouped into distinct types, although any cell in the same type
is slightly different.  Several distinct types of
organisms seem to be formed (even in a uni-cellular organism),
as may be called `species', although the present definition of species
by sexual separation may not
be applied.  Some of these types of states are stable against reproduction,
that is, a same type of unit is formed by the reproduction.  For example,
in the cell differentiation process, cells are diversified at the
initial stage to several types, and later are determined to 
keep the distinct cell types (i.e., to have recursiveness). Thus
(vi){\sl the origin of recursive states} is the next question to be answered.
Note that these types are not predetermined, and the variation within the type
can lead to further differentiation.  This differentiation process
is temporally organized in a tree-like structure, and sub-types with
smaller difference is also formed. Now (vii) {\sl the question of
hierarchical differentiation} is raised, as is seen in cell differentiation 
process, and possibly in the evolution.

Now let us come back to the question (iv), where
we note that the separation cannot be complete, 
since we live in interacting world.
Forming a good interface to cut the
individual from the outer world is the only possibility.
How is this possible?  Each individual unit must have information of the outer
world embedded within it.  The interface forms 
a mutual relation between the whole and individual;
the whole consists of individuals, but the property of an individual 
is also governed from the higher (ensemble) level of elements (whole).
This mutual feedback process, which is a key issue
in complex system studies, may be termed as a kind of
complementarity between the whole and part (individual).  
In the study of complex systems, we search for a mechanism
(iix){\sl how a global information on the ensemble of elements is embedded into
each individual one}.

When one considers a problem of biology, 
we have often seen complementarity between two groups:
Roughly speaking, one group is characterized by
symbol, rule, syntax, discreteness, and part, while the other by
pattern, behavior, semantics, continuity, and the whole.

The current trends in molecular biology aim to
explain the latter from the former.
In the study of development, one is interested in how a set of instructions 
given by digital information on DNA leads to the body of complex organisms.
On the other hand, such instruction itself should have been evolved 
through the history of life, as long as one does not assume the ``designer" of 
Life. In this context it is meaningful
to ask the reverse question mentioned at (iii), rephrased as; 
{\sl How is a set of syntactic rules 
formed from complex developmental process}? 

The complementarity between the above two groups
is often discussed in quantum-mechanical context.  Indeed
some believe that quantum mechanics is essential to 
biology.  However the concept of complementarity itself is
not necessarily associated with quantum mechanics, but is more general\cite{Bohr}.
In the study to be pursued here, we try to demonstrate that interacting
dynamics of internal states leads to such complementarity by allowing for
autonomous change of the rule of the dynamics itself.

Now let us come back to the problem (iv) on the formation of unit:
Units do not exist a priori.  These are formed
through interactions.  Since the behavior is diverse enough and
the separation of units from the outside is not complete, the units cannot be
completely identical.  For example, all cells are not identical
in contrast with an elementary particle in physics.
A biological system is essentially heterogeneous.
The importance of heterogeneity lies in the ability  to map the outer world 
into a unit, which again comes to the problem (iix) ({\sl embedding
global information into each individual unit}).  In the
development, cells, besides their distinct
types, have global information on their position, possibly 
represented in their modulation.
This is why we adopt dynamics with a continuous state later, instead of
discrete-state dynamics like cellular automata.

The hierarchy in a heterogeneous system is different from
that in a homogeneous system.  According to (iix), the information on
an ensemble can be embedded into the variation of
units of the same type, and the relation between the ensemble and unit is
bidirectional.  This hierarchy continues further.  A tissue 
consisting of cells, for example, is differentiated.  Now we have to answer
(ix){\sl how the formed, higher-level ensemble again acts
as a unit for diversification and differentiation}.
This higher-level unit works as a unit for reproduction.  Thus the last 
question to be addressed is (x) {\sl the origin of individuality, 
which reproduces itself at an ensemble level}.

Let us sum up the postulates that should be answered as
complex systems studies for biology.

\begin{itemize}

\item
(i) mechanism to create diversity

\item
(ii) the ability to form a higher level unit consisting of an ensemble of
the original units, stable as a state at each instant and also through
developmental process

\item
(iii) formation of a syntactic rule from complex mutual relationship

\item
(iv) formation of a unit to separate it from outside

\item
(v) formation of discrete states leading to (cell) types, and also
analogue modulation of the state

\item
(vi) recursivity of a state, preserved by the reproduction 

\item
(vii) hierarchy of differential process, characterized in the time 
course and also in the phenotype space

\item
(iix) mechanism to map the global information (on ensemble) to each individual
unit

\item
(ix) higher level of differentiation

\item
(x) formation of a higher-level reproduction unit

\end{itemize}

In the cell biology \cite{Cell}, the above list corresponds to (i) diversity of cells
(ii) formation of cell society, leading to multi-cellular organisms or tissues
where developmental and ongoing stability of the cell society is sustained
(iii) formation of a correspondence between genetic information and
phenotype, and later the emergence of cell differentiation rule 
(iv) origin of a cell with a membrane that suitably separates it from the
outside
(v) formation of discrete cell types with modulation within each type of cell
(vi) determined cell differentiation (i.e., formation of cell type keeping 
its type)
(vii) hierarchy of differentiated cell types, both in the developmental
course and in the difference of cell characters
(iix) modulation of cell character in each type of cell, reflecting on the
surrounding cells (or number distribution of cells of each type in the ensemble)
(ix) differentiation of an ensemble of cells (tissue), and
(x) ensemble of cells acting as the reproduction unit, in other words, the
reproduction of multi-cellular organism, and the formation of individuality.

It should be noted that (most of) the above problems are essential to
a biological system in general, not only to a cell society but also
to neural and immune networks, society of organisms, 
dynamical stability of ecosystem, evolution of language, and so forth.

\section{Intra-inter dynamics}

What type of model should one adopt to capture the above constraints?
In a biological system it is important
to capture the interplay between inter-unit and intra-unit dynamics \cite{CML}.
Such ``intra-inter dynamics" is essential to cell biology, where
complex metabolic reaction dynamics
in each unit (cell) is affected by the interaction among cells.
An ecological system also consists of interacting
units with internal dynamics. 
In neural systems also, the intra-inter dynamics is relevant
to the formation of internal images.

Another missing factor in conventional dynamical systems studies for 
modeling biology is the ``fluidity" of dynamical systems itself.
In a dynamical system approach, we have a set of variables to 
represent states, the rules of their temporal evolution 
and, initial and boundary conditions of them.  The rules and
initial and boundary conditions are given in advance and fixed
independently of the change of state variables.  
In biological problems, such independence
may not be valid.  For example, 
the number of variables itself changes with time, through, for example,
by cell divisions and cell deaths.
In developmental process, initial and boundary conditions of states
are chosen so that the reproduction continues, from their
mother's states.

In our approach we allow for ``dynamics of dynamics".
For example,  the change in the degrees of freedom is allowed, where
formation of
a unit acting as a ``partial system" is possible, which selects its initial 
and boundary conditions.

As a specific example of the scheme of intra-inter dynamics, let us
consider cell differentiation.
Here a set of chemical concentrations in a cell is chosen as the state 
variables. Internal dynamics consists of several biochemical reaction processes,
while there exists cell-to-cell interaction through diffusion of chemicals and
other signal transmission.  The change of dynamics
itself is due to the cell division and death depending on 
the cellular state, by which the number of
degrees of freedom varies.
We will see that a rule at a higher level for cell society is
formed from complex dynamical behavior arising from the lower level 
(biochemical) dynamics.

Now the remaining questions are the
choice of internal dynamics, interaction, and division process.
We have several possibilities in the choice, and  
the model can be classified according to
the complexity of the three processes; 
(a) the degrees of freedom in each cell and the form of the internal
dynamics,
(b) form of cell-to-cell interaction, and 
(c) meta-dynamical structure to change the cell numbers.

First, it may be interesting to reconsider previous models on cell biology 
from the present viewpoint.  
In the pattern-formation mechanism of
the Turing instability mechanism \cite{Turing},
the internal dynamics is not so well defined, although it 
provides the most important step
for the interaction-based approach in cell differentiation.
As for the level of internal dynamics there are two pioneering studies.
The importance of temporal oscillations in cellular dynamics was stressed
by Goodwin\cite{Goodwin} (see also \cite{Hess}).  
Diverse cell types were attributed to coexistence of many attractors
by Kauffman \cite{Kauff},
by adopting Boolean network dynamics (cellular-automaton type dynamics). 
Here cell-to-cell interaction is neglected, while some efforts to
include them are discussed following the Turing instability mechanism.
In these models, no meta-dynamical structure is included.

In these few years, we have studied several models of the
above intra-inter dynamics[8-13].  We choose a set of chemical variables
to assign each cellular state.  As for the internal dynamics 
we assume some oscillatory dynamics.  So far we have studied
;(see also Table I)

\begin{itemize}

\item
(A) phase dynamics with instability (for example by the circle map of
the phase $\theta_{n+1} =\theta_n +(K/2\pi)sin(2\pi \theta _n)$ \cite{KK-Rel,KK}

\item
(B) multi-phase dynamics given by a coupled circle map \cite{KK}

\item
(C) simple oscillatory dynamics with three chemicals\cite{KK-TY1}

\item
(D) oscillatory dynamics with several chemicals adopting
auto-catalytic reactions \cite{KK-TY2,KK-TY3}

\item
(E) chaotic dynamics with several chemicals adopting
auto-catalytic reactions \cite{CFKK}.

\end{itemize}

\vspace{.2in}

Table I

\begin{tabular}{|l|l|l|l|l|}\hline
 Model&Degrees of freedom & Internal  & Differentiation\\ \hline \hline
(A) & phase& circle map &   growth speed \\ \hline
(B) &multiple phases&coupled circle map &  chemical role  \\ \hline
(C) &phases and amplitudes&simple oscillation  & amplitude \\ \hline
(D) &phases and amplitudes& oscillation  & determined  \\ \hline
(E) &phases and amplitudes& chaos &  hierarchy, higher-level \\ \hline
\end{tabular}

\vspace{.2in}

In the models (C)-(E), 
there is a set of chemical variables, and
biochemical network within each cell, whose concentration changes
according to Michaelis-Mentens type catalytic reaction, where
chemicals catalyze each other, or themselves.

As for the interaction range, we have studied  global interaction with 
a homogeneous all-to-all coupling mostly, but have also
studied  local coupling among cells
within a given range.  
In the former case, cells are assumed to interact with each
other through the media.
For the interaction form, we have adopted
active transport process to get resources, for
(A)-(C), the diffusion process for (E), and both for (D).
The active transport here means that the ability to get chemicals from
the media depend on the chemicals in a cell.  Nutrition chemicals
are slowly poured into the media, and
the  interaction 
between cells leads to competition for nutrition chemical.

The cell division is assumed to occur when some 
chemical products (e.g., membrane or DNA) are accumulated beyond some
threshold, or a cell volume calculated from the chemicals within becomes
twice.  For both cases, the condition is of an
integral-type.  In the former case, the threshold
condition is given as an intergal of some chemicals
concentration, while for the latter case, 
the cell volume is expanded through the transport of chemicals into it,
until the cell divides into two.
For the `abstract' phase models of (A) and (B), the condition is not
such straightforward.  Here it is given by
the threshold for the accumulated number of rotation of phases,
since the rotation of phase in the model is brought about by the
flow of nutrition term in each cell.  For all models,
the concentrations of chemicals of two divided cells
are chosen to be almost identical upon the division.

The death condition is also determined by
a cellular state (e.g., if the total amount of chemicals is
less than some threshold, the cell dies).  The cell death process
usually sets in at a later 
stage of developmental process, where the competition for resources is
higher.

For all the models, diversification of cells and grouping of cells
by the clustering of phases of oscillators
are observed.  In the model (A), inactive cells without
division  and rapidly replicating cells are separated at some temporal regime.
In (B), roles of chemicals also differentiate, while the two
temporal regimes alternate, one with 
a society with diverse cells, and the other consisting of a homogeneous cells.
In (C), the amplitudes of cellular oscillations differentiate
to large and small ones, and the stability of an ensemble of cells is
found. In the present paper we generalize the results obtained
in the model (D) and (E).

Before describing the scenario of the 
differentiation extracted from the simulations, we note two 
backgrounds of the theory, one from dynamical systems and 
the other from experiments.

The scenario is based on the study of globally coupled chaotic 
systems \cite{KK-GCM}. In the study, 
tiny differences among the elements are amplified
through chaotic instability, which then leads to dynamical clustering of the
elements.  The temporal pattern of the clustering is robust against
external noise or is deterministic even if 
differences in the initial state are given stochastically.
Our model takes into account of the feature of the globally coupled chaotic 
system, although the internal dynamics in the model (D) is not necessarily
chaotic.  On the other hand, the change in the degree
of freedom is brought about by the cell division in the course of cell 
differentiation.  Hence the dimension of the phase space is no longer fixed, 
and the orbital instability there cannot be characterized by ordinary
dynamical systems, and we may encounter the situation what we call
open chaos \cite{KK-Rel,Alife}.  

On the other hand, stability at an ensemble level is studied
as collective dynamics in globally coupled maps \cite{KK-MF}.
By including meta-dynamics ( dynamics to change the parameter),
dynamic stability to keep the diversity is found, and the concept of
`homeochaos' is proposed \cite{homeo}.

Although our goal is the reinterpretation of the cell biology \cite{Cell}
from our intra-inter dynamics viewpoint,
it may be interesting to point out two specific
experimental results here.
Yomo and his colleagues reported that even under
single external condition, the cells differentiate to some distinct
physiological states \cite{Yomo}.  In their experiment, it was shown that one
cell type of E. coli resulted in  a population with several distinct
cell types after successive cultivation in a well-stirred liquid culture.
Even under
the same initial and external conditions, the cells autonomously
differentiated.     

Rubin \cite{Rubin} and his collaborators have shown that a cell line from 
mouse epigenetically transforms to different types of foci in size
under the same condition.  In addition, the frequency of transformation and
types of the transformed cells were shown to depend on the cell density and
the history of the cell culture.  This suggests that transformation or
differentiation of cells
is dynamically generated by inter-cellular interaction.  

\section{Isologous Diversification}

From several simulations of the model 
starting from a single cell initial condition, Yomo and the author
proposed the following ``isologous diversification theory", as
a general mechanism of spontaneous differentiation of replicating
biological units \cite{KK-TY1,KK-TY2,KK-TY3}, which is extended
in \cite{CFKK} by adopting a model class (E).   Starting from a
single unit, the following process of differentiation occurs.

{\bf (1)  Synchronous oscillations  of  identical units}

Up to some number of cells, all cells are identical as to the chemical
concentration, which oscillate coherently.
Accordingly, the cells divide almost simultaneously, and the number of cells 
is the power of two.  

{\bf (2)  Differentiation  of the phases of oscillations  of  internal
states.}

When the number of units exceeds some number, they lose identical and
coherent dynamics.  Although the number depends on the choice of network
and the parameters, the loss of synchrony generally appears in order to 
see continuous growth in cell numbers.  Then
cells separate into several groups whose phases of oscillations are
close.  
At this stage, only the phases of oscillations are different by cells, but  
the temporal averages of chemicals, measured over periods 
of oscillations, are almost identical.   
The behavior here is due to the clustering of phases studied 
in coupled  nonlinear oscillators.
As has been discussed \cite{KK-Rel,KK-TY1}, this temporal 
clustering provides
time sharing for resources: Cells can get chemical resource successively in 
order by the difference of phase of oscillations,
because the rate of the transport of resources into a cell
depends on the chemicals within.

{\bf (3)  Differentiation  of the amplitudes of internal  states. }

After some divisions of cells, 
differences in chemicals start to be fixed by cells.
The average chemical concentrations and their ratios differ by cells,
even after taking the temporal average
over periods. Thus the behavior of states is differentiated to some types.
The orbits of chemical dynamics lie
in a different phase space region by types of cells.

It is also interesting to note that the frequency of oscillations is
also differentiated.
One group of cells  oscillates and divides faster than the other group.
Hence
the differentiation of inherent time scales of cells emerges
spontaneously through cell divisions.   

{\bf (4)  Transfer  of the differentiated state to the  offsprings  by
reproduction. }

After fixed differentiation, chemical compositions
of each group are inherited by their daughter cells.  
Cell state represented by average chemical composition 
remains to be identical by division,
and thus the cells keep the ``recursivity" by divisions.
It is important to note that the chemical characters are ``inherited"
just through the initial conditions of chemical concentrations after the
division, although we have not explicitly imposed 
any specific mechanisms to keep the type.

The determination of a cell has occurred
at this stage, since daughters of one type of cells preserve the type. 
By reproduction, the initial condition of  units  is
determined  to give the next generation of units of the same  type.   
Thus a kind of memory is formed, attained through
the transfer of initial conditions on chemical concentrations by
the cell division.

The cellular memory at this fourth stage is formed as a result of the 
selection of
initial conditions for a cellular state (i.e., a partial system of the total 
dynamical system).  
One might think that this selection is nothing but a choice of basin of 
attractions
for a multiple attractor system.  If the interaction were neglected,
this would be basically correct.  In our case, this is not true.
Indeed most of the dynamical states of cell types do not exist as an
attractor but are stabilized through interaction.
The observed memory lies not solely in the internal states but
also in the interactions among the units.  

To see this intra-inter nature of the memory explicitly,
one effective method is the transplantation experiment.
Numerically, transplantation experiments are carried out
by choosing determined cells (obtained at the normal diffusion process)
and putting them into a different set of surrounding cells,
and making a cell society that could not appear through the normal course
of development.
 
When a determined cell is transplanted to another cell society
with different cell type distribution,
the offspring of the cell remain to be of the same type,
unless the cell-type distribution of the society is strongly biased
( i.e., the ensemble consisting of the same type of the cell as transplanted).
The cell memory is preserved mainly in each cell, 
but suitable cellular interactions are also necessary to keep the
stability.
Generally speaking, internal cellular memory is maintained as long as
the diversity is sustained.  
The achieved recursivity is understood as the choice of 
internal dynamics through cellular interactions.

{\bf (5)  Hierarchy of organized groups. }

As the cell number increases, further differentiation proceeds.
Each group of cells further differentiates
into two or more subgroups.   Some cells behave as
a stem cell to support further differentiated cells.

For example, six types ("0","1","2","1a",1b"1c") of cells are found 
in \cite{CFKK}.
The differentiation rule here is found to obey
($0\rightarrow 0$,1, or 2; $1\rightarrow 1$,1a,1b,or 1c; $2\rightarrow 2$,
$1a\rightarrow 1a$ , $1b\rightarrow 1b$, and  $1c\rightarrow 1c$,
in the normal course of differentiation starting from a single cell).
Hierarchical rule of differentiation is thus
generated.  Although the number of cell types and the
rule of differentiation depend on the choice of chemical networks,
generation of a hierarchical rule (written by the tree-type
diagram constructed from the above rule) is generally observed,
in a class of models (E).

Thus differentiations obey a specific rule, which is
given as a change of internal states but depends on
the dynamical interaction among cells.
The rules of differentiation and the higher level
dynamics emerge as the interaction of cells with internal dynamics.

Often the switching of cell types to further subgroups are
given by a stochastic automaton rule, where the stochasticity is
supported by the chaotic intra-cellular dynamics\cite{CFKK}.
The rate of the differentiation or the replication
( e.g., the choice which arrow from $0\rightarrow 0,1,2$ is selected)
depends on the cell-type distribution.  For example,
when some of type-1 cells are removed, the differentiation rate
$0 \rightarrow 1$ increases.  With this regulation mechanism to
recover the decreased cell type,
the stability of the distribution of cell types is kept.
The global stability of the whole system is thus obtained, by
spontaneous regulation of the rates of the differentiations.

{\bf (6) Formation of higher level dynamics and diversity of cell groups}

Since the rule of differentiation depends on the distribution of other
cell types, one can get an approximate dynamics for the
population of each cell type.  This is a higher level than
a cell type from chemicals, i.e., tissue from cell types.
This population-level dynamics is stochastic, since 
the information on the number of cell types
is not complete, where the lower-level information on the
internal state (of chemical concentrations) is discarded.
It is interesting to note that the macroscopic flow chart
on the number of cell types is formed in spite of the stochasticity.

In some models \cite{CFKK}, we have found that this
higher level dynamics allows for several stable states,
implying the coexistence of several stable cell distributions.
Indeed, in the models, we have found different sets of cell
distributions, starting from a single cell, depending on its
initial condition.  Thus we have found that the cell colony
is also diversified and differentiated to few groups.

\section{Discussion}

Let us discuss how the initial ten problems 
are resolved (or remain unsolved) in our intra-inter dynamics
approach.

\begin{itemize}

\item
(i) Diversification due to orbital instability in
each internal dynamics (open chaos)

In our study, the diversity is created through 
the orbital instability in dynamical systems. 
Chaotic {\sl attractor} is  not necessarily required,
since the transient dynamics with instability, in conjunction
with the change of degrees of freedom, is often sufficient
to provide the diversity.

As a corollary of this mechanism,
relevance of chemicals with low concentration is expected,
since tiny difference  of ``rare" chemicals 
between two cells is amplified to
lead to a macroscopic difference between them.
In our simulation, chemicals with tiny amounts in cells are relevant to the 
trigger to differentiations. It should be noted that
this relevance of chemicals with low concentrations is also supported
in physiological facts. Even at differentiated cells, difference is most 
remarkable for chemicals with low concentrations.  

\item
(ii) Stability of an ensemble level,
partly provided by collective dynamics of coupled
nonlinear elements

Stability at a macroscopic level has been discussed in dynamical 
systems theory, where dynamics of an ensemble of chaotic elements
keeps some stability through the interaction \cite{KK-MF,homeo}. 
In our simulation, removal of several cells of 
a given  type enhances the differentiation to the removed type, 
which restores the original ratio of types of cells.  
This macroscopic stability assures the robustness of developmental
process against perturbations including somatic mutations.

\item
(iii) Formation of a syntactic rule, partly by the rule for chaotic itinerancy;
(which, however,  waits for other mechanisms to be clarified)

The deepest question in this topic is the formation of semantic-syntactic
correspondence such as the phenotype-genotype one.  So far we have not 
succeeded in constructing a coherent scenario for it
within our intra-inter dynamics,
although in a model of a class (E), differentiation of 
chemicals between slow, inactive, and controlling variables and
fast, and controlled variables with chaotic dynamics are found, which may
support the postulate by corresponding the former to the role of
genotype and the latter to the phenotype\cite{KKTYtp}.

The problem we have addressed in the present paper is much simpler; the
formation of differentiation rule.  As is discussed later at (v),
each cell type is represented as a state in the phase space of the internal 
dynamics.  The transition rule is given by the switching among these states,
through interactions.  A related mechanism for the formation of
the switching rule has been discussed in chaotic itinerancy 
\cite{KK-GCM,Ikeda,Tsuda}.  It
is long-term dynamics with itinerancy over several
states through chaotic dynamics, often found in
high-dimensional dynamical systems. 
For our switching rule, we do not have clear 
dynamical systems representation, since
it is also interaction dependent, although the understanding of chaotic 
itinerancy\cite{Milnor}  may give a step towards the rule formation.

\item
(iv) Formation of a unit to separate it from outside: (not yet discussed)

Since we have assumed the existence of a cell,
separated out by a membrane, this question cannot be discussed as yet,
although it is essential when one considers the origin of life \cite{Tile}.

\item
(v) Formation of discrete (cell) types, first provided
by clusterings, and then by attracting states stabilized by the interaction

After the diversification of cell states, they are grouped
into several types.  Such grouping of oscillators by their
phases is first studied as clustering in coupled map system at the stage
(2).  Then the cells split into several groups whose orbits lie
in different parts in the phase space.
Here these states are represented as attracting states
stabilized by the interaction.  Only a discrete set of states 
exists, stable against perturbations, cell divisions,
and the variation of interaction in the natural course of
differentiation.

Difference of the phases of oscillations by the clustering
is given by ``analogue" means, and cellular states
at any phase of oscillations can exist in principle.
On the other hand, the differentiation based on the 
phase space position is digital, in the sense that only discrete levels
are allowed.  
Hence there are two levels of differences by cells, one for the
change of phases of oscillations, and the other for
the fixed differentiation.  

Note that the differences by phases of oscillations are not rigid, 
since the phase is easily diffused by external disturbances:  
Perturbations brought about by division are enough to shift 
the phase and destroy the memory of the previous clustering.  On the other 
hand, the ``digital" difference by the amplitude of oscillations is more 
rigid, since it is not shifted continuously as in the case of phase.  
This emergence of digital information
is the basis of the cellular memory at (vi).

\item
(vi) Recursivity of a state, preserved by the reproduction,
as a choice of initial conditions with digitalization of 
differentiated states

The recursive character of states is attained as the transfer of
initial conditions by cell divisions, so that the dynamics later
remains to be attracted to the same type.  Such attraction is
supported by the interaction.  When the cellular state is disturbed,
the intra-inter dynamics works to restore the orbit to have the recursivity.
The ``digital" distinction of chemical characters, that has emerged at (v), 
is relevant to preservation of the characters to daughter cells, since
analogue differences of phases may  easily be disturbed by the division 
process, and cannot be transmitted to daughter cells robustly.

In biological terms, the recursivity corresponds to determination
of differentiation. Our picture implies that there are levels of 
robustness of this determination.  Since the stability of states is given as 
a balance between
internal dynamics and the interaction, the robustness against the
external change is larger as the internal dynamics plays more role for the
determination of the state.

\item
(vii) Hierarchy of differential process, supported possibly
by hierarchical structure in a coupled chaotic system

We have observed the hierarchical differentiation, as
a successively smaller structure in the phase space\cite{CFKK}.
We have not yet clarified the condition to have such hierarchical
structure in the phase space, but it should be noted that (transient)
chaotic dynamics often provides a hierarchical structure,
as is studied in globally coupled maps\cite{KK-GCM}.

\item
(iix) Mechanism to map the global information to each individual
unit, provided by the interaction-induced modulation 

As mentioned, two types of memory coexist, analogue and digital ones.
The former gives information on the cell society, i.e.,
the distribution of cell types, while the latter
gives a distinct internal state on cell differentiation.  Indeed
the orbits of chemicals of each cell type are shifted slightly 
with the change of
the distribution of cells surrounding it.  Gobal information
on the cell distribution is embedded into this `analogue' change,
which modifies  the rate of differentiation.
Now one can see how the complementarity
in \S 1 between analogue/whole and digital/part is given by
the intra-inter dynamics approach.

We believe that such dual memory structure is a general feature in a 
biological system,
that exists as an interface between external environment and the
internal dynamics.  In cell biology, the ``analogue" difference
reflecting on the interaction is known as modulation \cite{Cell}.
According to our theory, this modulation determines the rate
of differentiation, and leads to the stability of cell society.

\item
(ix) Higher level of differentiation by the formation of higher-level
dynamics supporting the instability and several stable states

According to the mechanism (vii), there appears a higher level
dynamics for the population of cell types.  This higher-level dynamics
can again have instability to provide the diversity and the stability
to attract several states.  Although a few examples 
in model class (E) support this differentiation of cell
society, the condition to form this higher level is not
clear yet.

\item
(x) Formation of a higher-level reproduction unit by forming an interface
structure to separate an ensemble of units from other units

The origin of a multicellular organism, for example,
is directly related with our scenario.
For its origin, some mechanism of differentiation of 
identical cells is necessary which leads to divisions of labor.
In our theory, this mechanism is naturally explained.
Such differentiation generally appears, for cells to increase their number
within limited resources. Indeed,
the number of cells stops at a small level in our simulation,
when the internal chemical reaction network does not allow
for the clustering of oscillations and the state differentiation.

However, this mechanism is not sufficient for the origin of a multi-cellular 
organism. The ensemble of cells as a whole should reproduce as a recursive
unit.  To
study this origin of a higher level unit, we have studied a spatial
version of our models (A) and (D).  Instead of global interaction, we have 
chosen local interaction only within a range, and also added the mobility of 
cells according to the inter-cellular force depending on cellular states.
When the ranges for chemical interactions and for
inter-cellular force are of comparable order, the cells 
continue dividing and spread over the
space, by forming a spatially localized set of cells,
acting as a unit for reproduction\cite{KK-sp}.

\end{itemize}

Summing up, we have proposed intra-inter dynamics
to provide a new viewpoint to cell biology.
We believe that the present theory can generally be applied to
a variety of biological systems, because it
is based on our study of coupled dynamical systems,
which is expected to be universal in a class of interacting,
reproducing, and oscillatory units.
So far we believe that the proposed theory 
captures the essence of cell differentiation,
such as the loss of totipotency, origin of stem cells, differences in
growth rates, and developmental stability, and so forth.
Apoptosis and tumor formation are also discussed in the present 
interaction-based picture \cite{KK-TY3}. Indeed, the
``tumor"-like cell is found in the model (D)\cite{KK-TY3}, as is 
characterized by extraordinary differentiation, loss of internal chemical 
diversity, faster growth, and destruction of the diversity of cells.
Indeed formation of ``selfish" cells destroying the
diverse cell society is expected,
when a suitable condition of the interaction is lost,
since our differentiation process is not
programmed explicitly as a rule but occurs through the interaction
in our theory.

{\sl acknowledgments}

The present paper is based on the studies in collaboration  with T. Yomo and
C. Furusawa. I am grateful to  them
for stimulating discussions.  I would also like to thank S. Sasa and 
T. Ikegami, for discussions and Y. Oono for criticisms.
The work is partially supported by
Grant-in-Aids for Scientific
Research from the Ministry of Education, Science, and Culture
of Japan.

\end{document}